\documentstyle[preprint,tighten,aps]{revtex}

\begin{document}
\draft
\preprint{\parbox[t]{2 in}{CMU-HEP93-23\\
(DOE-ER/40682-48)}}

\title{Isoscalar-isovector mass splittings in excited mesons}
\author{Paul Geiger}
\address{Physics Department, Carnegie Mellon University, Pittsburgh,
PA 15213\\}
\maketitle

\begin{abstract}
  Mass splittings between the isovector and isoscalar
members of meson nonets arise in part from hadronic
loop diagrams which violate the Okubo-Zweig-Iizuka rule.
  Using a model for these loop processes which works
qualitatively well in the established nonets,
I tabulate predictions for the splittings and
associated isoscalar mixing angles
in the remaining nonets below about 2.5 GeV,
and explain some of their systematic features.
  The results for excited vector mesons compare
favorably with experiment.
\end{abstract}


\narrowtext

     The established meson nonets, with the exception of
the pseudoscalars, exhibit two notable regularities:
(i) the isoscalar members (which we shall generically denote by
${\cal F}$ and ${\cal F}^{\prime}$) are almost ideally mixed:
${\cal F} \approx {u\bar{u}+d\bar{d} \over \sqrt{2}}$ and
${\cal F}^{\prime} \approx s\bar{s}$, and
(ii) the ${\cal F}$ is nearly degenerate with its isovector
partner, ${\cal A}$.
    Though the $U(1)$ anomaly spoils these two rules of
thumb in the $0^{-+}$ sector, they hold well enough in
$1^{--}$, $1^{+-}$, $1^{++}$, $2^{++}$, $3^{--}$ and $4^{++}$
mesons that one may easily forget they are not
underwritten by any firm theoretical considerations.
    For while we should expect violations of (i) and (ii)
to be suppressed because they
entail violations of the Okubo-Zweig-Iizuka (OZI) rule \cite{OZI}
   (the ${\cal F-A}$ splitting is proportional to the
amplitude $A$ for $u\bar{u} \leftrightarrow d\bar{d}$ mixing
and the ${\cal F-F}^{\prime}$ mixing angle is proportional to
the amplitude $A^{\prime}$ for $\{u\bar{u}\,\,{\rm or}\,\,d\bar{d}\}
\leftrightarrow s\bar{s}$ mixing),
   the observed {\it degree} of suppression does not follow from
any general argument.
   In particular, in the large-$N_c$ expansion $A$ and
$A^{\prime}$ are proportional to
$N_c^{-1}$, the same as a typical OZI-allowed hadronic width,
leading one to expect mass splittings
$M_{\cal F}-M_{\cal A}$ of order $100$~MeV and
mixing angles $\theta-\theta_{\rm ideal}$ of order
$\tan^{-1}(1/2)$.

   In fact, Fig.~1 suggests a specific mechanism for
generating substantial $A$ and $A^{\prime}$:
   even if we suppose that the ``pure
annihilation'' time-ordering of Fig.~1(a) is small,
it seems difficult to arrange a suppression of the hadronic
loop diagram in Fig.~1(b) since the vertices there are OZI-allowed
and the loop momentum runs up to $\sim \Lambda_{\rm QCD}$
before it is cut off by the meson wavefunctions.
   This instance of a ``higher-order paradox''~\cite{LipkinOZIstuff}
was investigated in detail in Refs.~\cite{GeiIsg1,GeiIsg2}.
   There it was found that,
while individual intermediate states (such as $\pi\pi$,
$\omega\pi$, {\it etc.}) do indeed each contribute
$\sim 100$~MeV to $A$ and $A^{\prime}$, the sum over
{\it all} such states tends to give a much smaller net result,
of order 10~MeV in most nonets. In essence, this occurs because the
constituent quark and antiquark created
at the lower vertex of Fig.1(b) emerge in a
dominantly $^3P_0$ relative wavefunction
(a result inferred
from meson decay phenomenology~\cite{3P0refs,KokIsg}) and
the sum over intermediate states turns
out to closely approximate a closure sum in which
the created quarks retain their original quantum numbers,
hence they have no overlap with the final state meson
in all nonets except $0^{++}$.

   An obvious corollary to this explanation is that
properties (i) and (ii) may break down appreciably
for scalar mesons; this scenario was examined in Ref.~\cite{GeiIsg2}.
   However, smaller deviations can also be expected
in other nonets, simply because the cancellations among the
loop diagrams are not always perfect. In fact, as we will see,
the cancellations are {\it expected} to be less complete for some
excited nonets.
   In this paper I present predictions for the loop-induced
${\cal F-A}$ splittings and ${\cal F-F}^{\prime}$ mixing angles
in excited meson nonets, and explain some
systematics of these predictions. The excited vector
mesons are particularly interesting, as
the available experimental data indicates
a sizable ${\cal F-A}$ splitting in both the $2^3S_1$
and $^3D_1$ sectors.

\vskip 0.4 in
    The mixing amplitude of Fig.~1(b) is given by
\begin{equation}
  A(E) = \sum_{n} { \langle d\bar{d} \vert H_{pc}^{u\bar{u}} \vert
  n\rangle\langle n\vert H_{pc}^{d\bar{d}}\vert u\bar{u} \rangle
  \over  (E - E_n) },
  \label{eq:AofE}
\end{equation}
where the sum is over a complete set of two-meson intermediate
states $\{ \vert n \rangle \}$, and $H_{pc}^{f\bar f}$ is a
quark pair creation operator for the flavor $f$. Similar
formulas of course apply for $A^{\prime}(E)$ and
$A^{\prime\prime}(E)$ (the latter denotes the amplitude for
$s\bar{s} \leftrightarrow s\bar{s}$ mixing).
  The $^3P_0$ decay model leads to the following expression
for the 3-meson vertices:
\begin{eqnarray}
  \langle A \vert H_{pc} \vert BC \rangle&=&{2\over (2\pi)^{3/2}}
  \gamma_0 \phi
  {\bf \Sigma} \cdot \int d^3k \, d^3p \, d^3p^\prime
  \Psi({\bf p},{\bf p}^\prime)
  \Phi^*_B({\bf k}+{{\bf p}^\prime\over 2})
  \Phi^*_C({\bf k}-{{\bf p}^\prime\over 2})
  \nonumber \\
  &&\times \> ({\bf k}+{{\bf q}\over 2})
  \exp[{-{2r^2_q\over 3}({\bf k}+{{\bf q}\over 2})^2}] \;
  \Phi_A ({\bf k}-{{\bf q}\over 2}-{\bf p}).
  \label{eq:Our3P0}
\end{eqnarray}
Here the $\Phi$'s are meson wavefunctions (which out
of computational necessity we take to be harmonic oscillator
wavefunctions), while $\phi$
and ${\bf \Sigma}$ are flavor and spin overlaps, respectively.
   The matrix element is evaluated in the rest frame of the initial
meson $A$ so that ${\bf P}_B = -{\bf P}_C \equiv {\bf q}$.
   The intrinsic pair creation strength, $\gamma_0$,
and the ``constituent quark radius,'' $r_q$, are
parameters which we fit to measured decay widths.
   The function $\Psi ({\bf p},{\bf p}^{\prime})$
contains a flux-overlap factor that arises in
the flux-tube breaking model~\cite{KokIsg}, and also a
``color-transparency'' factor, which incorporates a
reduction in the pair creation amplitude
when the quark and antiquark in meson $A$ are close
enough to screen each other's color charges~\cite{Parms}.
   The intermediate states are labeled by the oscillator
quantum numbers $\{n,\ell,m, S, S_z\}$ of mesons B and C, plus the
momentum and angular momentum of the relative coordinate.
   Most of the contribution to $A(E)$ comes from low-lying states
($\ell_B,\ell_C \lesssim 3$ and $n_B,n_C \lesssim 1$) but
we sum up to $\ell_B,\ell_C \approx 8$ and $n_B,n_C \approx 4$
in order to see good convergence.

   By writing the mixing amplitudes as contributions to meson mass
matrices,
\begin{equation}
  M = \left[ \matrix{
      m+A     &       A      &   A^{\prime}                   \cr
       A      &      m+A     &   A^{\prime}                   \cr
  A^{\prime}  &  A^{\prime}  & m+\Delta m + A^{\prime\prime}  \cr
  }\right]~~~,
  \label{eq:MassMatrix1}
\end{equation}
in the $\{u\bar{u},d\bar{d},s\bar{s}\}$ basis, or
\begin{equation}
  M = \left[ \matrix{
   m  &         0          &       0                       \cr
   0  &        m+2A        &  \sqrt{2}A^{\prime}           \cr
   0  & \sqrt{2}A^{\prime} & m+\Delta m + A^{\prime\prime} \cr
  }\right]~~.
  \label{eq:MassMatrix2}
\end{equation}
in the $\{ {(u\bar{u}-d\bar{d}) \over \sqrt{2}}, \;
{(u\bar{u}+d\bar{d}) \over \sqrt{2}},\; s\bar{s} \}$ basis,
one sees that, when the mixings are small,
$2A$ is the $\cal F-A$ mass difference
and $\sqrt{2}A^{\prime}/ \Delta m$ is the ${\cal F-F}^\prime$
mixing angle.

   Equation~(\ref{eq:Our3P0}) and our techniques for computing it,
our procedure for fitting its parameters, and our sensitivity
to those parameters, are discussed in detail in
Refs.~\cite{GeiIsg1,GeiIsg2}.

\vskip 0.4 in
   Table~\ref{tab:Results} contains our results.
   The $^3S_1$, $^1P_1$, $^3P_1$, $^3P_2$, $^3D_3$,
and $^3F_4$ nonets were already analyzed in Ref.~\cite{GeiIsg2};
we include them here to illustrate the level of accuracy
that may be expected for the new predictions~\cite{DiagCaveat}.
   (Note that the rather poor $^3P_2$ prediction is by far the most
sensitive to parameter changes -- it moves from $-3$~MeV to
$-38$~MeV when $\beta$ is changed from $0.40$~GeV to
$0.45$~GeV.)
   By also taking into account the stability of our results
under parameter variations, as well as the observed accuracy of the decay
analysis in Ref.~\cite{KokIsg}, we estimate that our mass-splitting
predictions are reliable up to a factor of two or so.

   While the mass splittings mainly measure $A$, mixing angles
give information about $A^{\prime}$. The mixing angles in
column 3 are defined by
\begin{eqnarray}
 {\cal F}          &=&  {\left\vert u\bar{u}+d\bar{d} \over \sqrt{2}
\right\rangle}
                  \cos\phi - \left\vert s\bar{s}\right\rangle \sin\phi
\nonumber \\
 {\cal F}^{\,\prime}&=& {\left\vert u\bar{u}+d\bar{d} \over \sqrt{2}
\right\rangle}
                  \sin\phi + \left\vert s\bar{s}\right\rangle \cos\phi \; .
\label{eq:MixingAngles}
\end{eqnarray}
Note that $\phi = \theta - \theta_{\rm ideal}$, where $\theta$
is the octet-singlet mixing angle and
$\theta_{\rm ideal} = \tan^{-1}(1/\sqrt{2})$.
   We do not list experimental results for mixing angles
since their extraction from meson masses is very model
dependent~\cite{PDG}, and though they
are measured quite directly by decay branching ratios,
these ratios are known only poorly for the interesting
({\it i.e.}, substantially mixed) $^1P_1$ and $^3P_1$ states.
   As with the mass splittings, our values
for $\theta-\theta_{\rm ideal}$ should only
be trusted to within a factor of about two.
   Such large theoretical uncertainties of
course mean that Table~\ref{tab:Results}
can only be taken as a rough guide. Nevertheless, it clearly
contains some definite qualitative predictions.
   Most interesting are the excited vectors, $2^3S_1$ and
$^3D_1$; some properties of these states have been extracted
from $e^+e^-$ experiments~\cite{PDG,DonnVectors}, so we
can compare with our calculations.
   For the radial excitations, we find
$m_{\omega^\prime} - m_{\rho^\prime} = -53$~MeV
and $\phi=-26^{\circ}$.
Most of the splitting here comes from $A^{\prime}$ rather than $A$,
{\it i.e.}, just as with the $^3P_1$ nonet, strange
intermediate states (in particular $K^*\bar{K} + \bar{K^*}K$)
are the source of most of the OZI-violation.
  The predicted mixing angle
is quite large but has only moderate effects on the
branching ratios to non-strange final states,
since the flavor overlaps for such decays are proportional
to $\cos^2\phi$.
   Thus, for example, we find that
${\Gamma(\omega^\prime \rightarrow   \rho\pi) \over
 \Gamma(\rho^\prime   \rightarrow \omega\pi)}$ is reduced
from 3 to 2.4  by the flavor overlap factor (and suffers a
further reduction to about 1.9 due to the decreased
phase space of the $\omega^{\prime}$).
  On the other hand, the flavor-overlap part of
${\Gamma(\omega^\prime \rightarrow K^*\bar{K}) \over
 \Gamma(\omega^\prime \rightarrow K  \bar{K})}$, which
goes like
$\left\vert { \cos\phi-\sqrt{2}sin\phi \over
\cos\phi+\sqrt{2}sin\phi }\right\vert ^2$, is enhanced
by a factor of almost 30 over the ideal-mixing
prediction. The flux-tube breaking model
of Ref.~\cite{KokIsg} (which is probably reliable to within
a factor of 2) predicts
$\Gamma(\omega^\prime \rightarrow K^*\bar{K}) \approx 20$~MeV and
$\Gamma(\omega^\prime \rightarrow K  \bar{K}) \approx 30$~MeV
for an ideally mixed $\omega^{\prime}$; with our mixing angle of
$-26^{\circ}$ the predictions become approximately
$40$~MeV and $2$~MeV, respectively~\cite{ImAprime}.

   The mixing amplitudes in the $^3D_1$ nonet
are unusually large: $A \approx -130$~MeV and
$A^{\prime} \approx -160$~MeV, thus our perturbative
calculation is probably less
trustworthy here than in other nonets. Nevertheless we
have significant qualitative agreement with the
experimental $\omega^{\prime\prime} - \rho^{\prime\prime}$
splitting of $(-106 \pm 23)$~MeV -- the
largest measured splitting in Table~\ref{tab:Results}.
   The biggest individual contributions to $A$ in this
sector come from the $a_0 \, \rho^{\prime\prime}$ and
$a_1 \, \rho^{\prime\prime}$ intermediate states.
   The phenomenology of the large negative $^3D_1$
mixing angle is very similar to the $2^3S_1$ case:
${\Gamma(\omega^{\prime\prime} \rightarrow   \rho\pi) \over
 \Gamma(\rho^{\prime\prime}    \rightarrow \omega\pi)}$ is reduced
to about 2, and the predictions of Ref.~\cite{KokIsg},
$\Gamma(\omega^{\prime\prime} \rightarrow K^*\bar{K}) \approx 10$~MeV and
$\Gamma(\omega^{\prime\prime} \rightarrow K  \bar{K}) \approx 40$~MeV
become approximately $20$~MeV and $3$~MeV, respectively.

   A final comment on the results of Table~\ref{tab:Results}
concerns the average magnitudes of $M_{\cal F}-M_{\cal A}$
and $\theta-\theta_{\rm ideal}$ in the various nonets.
   Confining attention to the radial ground states, the
apparent trend is for the mixings to start out small
in the S-wave mesons, become considerably larger in the P and D
wave nonets and then decrease again for the
F and G (and higher) nonets.
   This pattern can be understood as follows.
   It was shown in Ref.~\cite{GeiIsg1} that
for a particular choice of the pair creation form
factor, the closure sum corresponding to Eq.~(\ref{eq:AofE})
can be written as a power series in a variable
$\lambda$ which is a function only of $\beta$ and $r_q$.
   The coefficient of the $\lambda^k$ term is the sum
of all loop graphs whose intermediate states satisfy
$2(n_b+n_c)+(\ell_b +\ell_c +\ell_{\rm rel}) = k.$
   Since the closure sum vanishes for any $\lambda$, it follows
that each subset of graphs corresponding to a particular value
of $2(n_b+n_c)+(\ell_b +\ell_c +\ell_{\rm rel}) \equiv 2N+L$
sums to zero.
   (For example, in the $N=0$ sector, intermediate
states containing two S-wave mesons in a relative P-wave
exactly cancel with intermediate states where an S-wave meson
and a P-wave meson are in a relative S-wave.)
   The intermediate mesons in each subset have similar masses,
hence the cancellation tends to persist in the full
calculation with energy denominators.
   With P- and D-wave initial states,
the terms with $(2N+L={\rm constant})$ no longer exactly cancel;
some of the $(2N+L= {\rm constant}+2)$ terms must be
added~\cite{EvenOdd}, and
the significant mass splittings among such states tends to
spoil the cancellations when energy denominators are inserted.
    For F- and G-wave initial states the cancellation requires
the $(2N+L= {\rm constant})$, $(2N+L= {\rm constant}+2)$, and
$(2N+L= {\rm constant}+4)$ terms, so one might expect even
worse deviations from the closure result. However, some of these
terms vanish identically because the highly excited initial
state does not couple to them: in the $^3P_0$ model, the orbital
angular momentum of the initial state, $\ell_A$, can differ from
$L$ by at most one unit, thus (for example) F-wave mesons do not
couple at all to intermediate states which have $L=0$.
   Hyperfine splittings in the $L=0$ sector cause the largest
deviations from closure (note also that deviations due to radial and
orbital splittings are largest among low-lying intermediate states),
thus by decoupling from $L=0$ the F- and G-wave mesons end up experiencing
less OZI-violation than P- and D-wave mesons.

\vskip 0.4in
   In summary, Table~\ref{tab:Results} ought to provide a useful
rough guide to isoscalar-isovector mass splittings and mixing
angles in excited meson nonets. There is no good reason to expect
$\left\vert M_{\cal F}-M_{\cal A}\right\vert \, \lesssim \, 10$~MeV
in general. In fact, it is probable that the splittings in P- and D-wave
mesons, as well as in radial excitations, will be substantial.

\vskip 0.4in
  I thank Nathan Isgur for discussions.
This research has been supported, in part, by an NSERC of Canada
fellowship and by the U.S. Dept. of Energy under grant
No. DE-FG02-91ER40682.

\begin{figure}
  \setlength{\unitlength}{1.2 mm}
  \begin{picture}(120,140)(0,-10)
    \thicklines
    \put(90,10){\oval(10,40)[tr]}  \put(110,10){\oval(10,40)[tl]}
    \put(90,50){\oval(10,40)[bl]}  \put(110,50){\oval(10,40)[br]}
    \put(100,50){\oval(10,30)[bl]} \put(100,50){\oval(10,30)[br]}
    \put(90,100){\oval(10,40)[br]} \put(110,100){\oval(10,40)[bl]}
    \put(90,60){\oval(10,40)[tl]}  \put(110,60){\oval(10,40)[tr]}
    \put(100,60){\oval(10,30)[tl]} \put(100,60){\oval(10,30)[tr]}
    \put(85,60){\line(1,-1){10}}   \put(115,60){\line(-1,-1){10}}
    \put(85,50){\line(1, 1){4}}    \put(95,60){\line(-1,-1){4}}
    \put(105,60){\line(1,-1){4}}   \put(115,50){\line(-1,1){4}}
    \put(94,5){$u$}                \put(104,5){$\bar{u}$}
    \put(94,103){$\bar{d}$}        \put(104,103){$d$}

    \put(30,10){\oval(10,80)[tl]}  \put(30,10){\oval(10,80)[tr]}
    \put(30,100){\oval(10,80)[bl]} \put(30,100){\oval(10,80)[br]}
    \put(24,5){$u$}                \put(34,5){$\bar{u}$}
    \put(24,103){$\bar{d}$}        \put(34,103){$d$}

    \put(5,90){\LARGE (a)}         \put(70,90){\LARGE (b)}
  \end{picture}
  \caption[x]{(a) OZI-violation by ``pure
   annihilation.'' (b) A different time ordering of the same
   Feynman graph gives OZI-violation via two OZI-allowed
   amplitudes. (In both cases, time flows upward.)}
\end{figure}
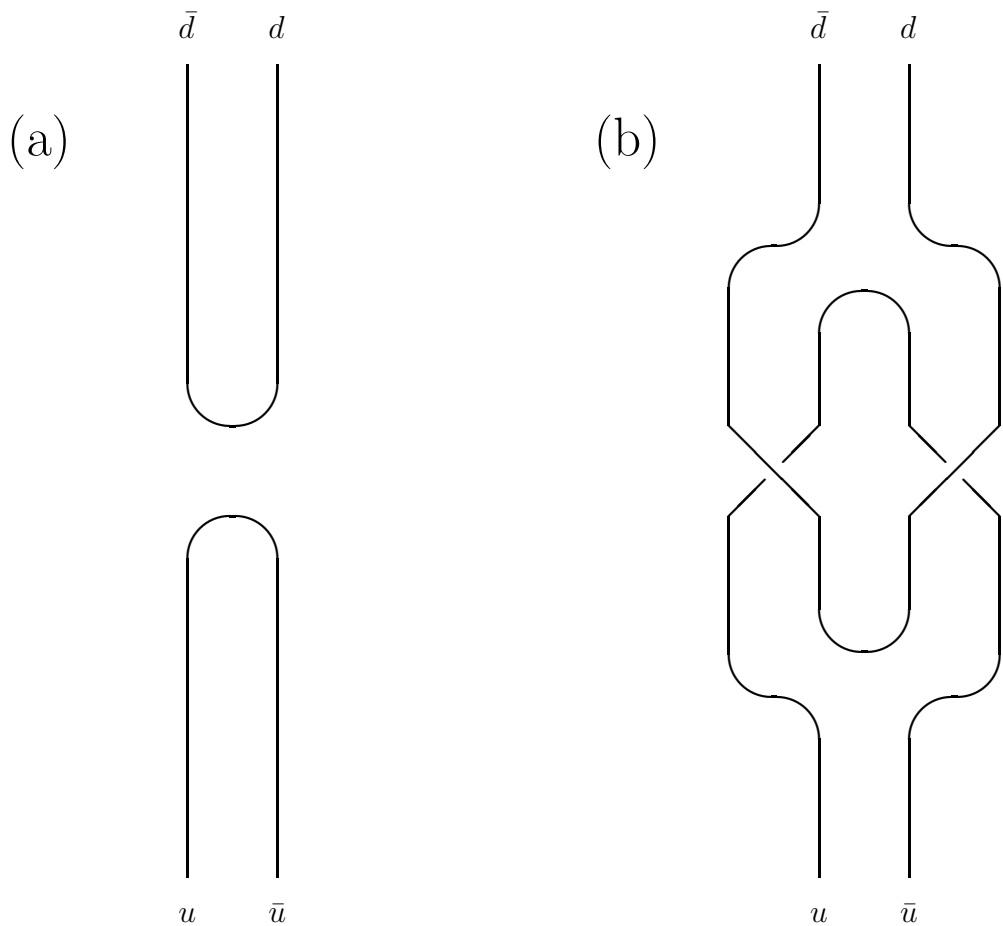

\widetext
\begin{table}
\caption{Mass splittings and mixing angles from hadronic loops.
The measured values are taken from Ref.~\protect\cite{PDG}.}
\begin{tabular}{cccc}
       & Predicted                       & Measured  &  Predicted \\
 Nonet & $M_{\cal F}-M_{\cal A}\;$ (MeV) & $M_{\cal F}-M_{\cal A}\;$ (MeV)
       & $\theta-\theta_{\rm ideal}\;$(degrees) \\
\tableline
$  ^3S_1 $ & $  4$  &  $ 14 \pm  2$   & $ -1$  \\
$ 2^3S_1 $ & $-53$  &  $-71 \pm 30$   & $-26$  \\
$ 3^3S_1 $ & $-51$  &                 & $ -3$  \\
           &        &                 &        \\
$  ^1P_1 $ & $-63$  &  $-64 \pm 24$   & $-15$  \\
$  ^3P_1 $ & $-18$  &  $ 22 \pm 30$   & $ 24$  \\
$  ^3P_2 $ & $ -3$ to $-38$\tablenote{See text.}
  &  $-44 \pm  6$   & $ -7$  \\
$ 2^1P_1 $ & $ 42$  &                 & $ 11$  \\
$ 2^3P_1 $ & $-48$  &                 & $ -7$  \\
$ 2^3P_2 $ & $ 26$  &                 & $-12$  \\
           &        &                 &        \\
$  ^1D_2 $ & $-48$  &                 &  $-1$  \\
$  ^3D_1 $ & $\approx -200^{\rm a}$
                    &  $-106\pm 23$   &  $\approx -25^{\rm a}$ \\
$  ^3D_2 $ & $ 32$  &                 &  $ 7$  \\
$  ^3D_3 $ & $  6$  &  $-24 \pm  8$   &  $ 1$  \\
$ 2^1D_2 $ & $ 48$  &                 &  $ 3$  \\
$ 2^3D_1 $ &$-121$  &                 &  $-6$  \\
$ 2^3D_2 $ & $-47$  &                 &  $-4$  \\
$ 2^3D_3 $ & $-29$  &                 &  $-4$  \\
           &        &                 &        \\
$  ^1F_3 $ & $ -2$  &                 &  $ 5$  \\
$  ^3F_2 $ & $ -2$  &                 &  $ 0$  \\
$  ^3F_3 $ & $-21$  &                 &  $-7$  \\
$  ^3F_4 $ & $ 28$  &  $12 \pm 36$    &  $-1$  \\
           &        &                 &        \\
$  ^1G_4 $ & $ -1$  &                 &  $ 2$  \\
$  ^3G_3 $ & $-14$  &                 &  $ 1$  \\
$  ^3G_4 $ & $ -2$  &                 &  $-3$  \\
$  ^3G_5 $ & $-14$  &                 &  $-1$  \\
\end{tabular}
\label{tab:Results}
\end{table}


\begin{references}
\bibitem{Parms}
The parameters used in the present calculation are the
same as those used in (the third column Table I of)
Ref.~\cite{GeiIsg2}:
$r_q=0.25$~fm, $\beta=0.40$~GeV, $b=0.18$~GeV$^2$,
and $t=g=1$.
\bibitem{DiagCaveat}
Note that in \cite{GeiIsg2} we simply reported the $A$
and $A^{\prime}$ amplitudes while in this paper we have
inserted them into the mass matrices and diagonalized.
  In practice this has negligible effects on the numerical
results, except in the $^3P_1$ and $2^3S_1$ nonets, where
$\vert A^{\prime} \vert$ is significantly larger than
$\vert A \vert$ so that most of the mass splitting
actually comes from $A^{\prime}$.
\bibitem{ImAprime}
Note that we have calculated only the real parts of
the mixing amplitudes; a significant imaginary component
in $A^{\prime}$ would alter these branching
ratio predictions.
\bibitem{EvenOdd}
Note that parity requires $L$ to be even (odd) whenever
$\ell_A$ is odd (even).
\end{references}
\end{document}